\documentclass[conference]{IEEEtran}
\IEEEoverridecommandlockouts
\usepackage{cite}
\usepackage{amsmath,amssymb,amsfonts}
\usepackage{xfrac}
\usepackage{bm}
\usepackage{algorithmic}
\usepackage{algorithm}
\usepackage{graphicx}
\usepackage{wrapfig}
\usepackage{textcomp}
\usepackage{xcolor}
\usepackage{pgf}
\usepackage{lmodern}
\usepackage{tikz}
\usetikzlibrary{arrows.meta, positioning}
\usepackage{hyperref}
\def\BibTeX{{\rm B\kern-.05em{\sc i\kern-.025em b}\kern-.08em
    T\kern-.1667em\lower.7ex\hbox{E}\kern-.125emX}}

\everymath=\expandafter{\the\everymath\displaystyle}

\usepackage{siunitx}
\begin{document}

\title{Momentum-Accelerated Online Feedback Optimization for Power System Flexibility}

\author{
    \IEEEauthorblockN{
        Florian Klein-Helmkamp\IEEEauthorrefmark{1}, Matthis Berger \IEEEauthorrefmark{1}, Irina Zettl \IEEEauthorrefmark{1},
        and Andreas Ulbig\IEEEauthorrefmark{1}\IEEEauthorrefmark{2}
    }
    \IEEEauthorblockA{
        \IEEEauthorrefmark{1}IAEW at RWTH Aachen University, Aachen, Germany\\
        \IEEEauthorrefmark{2}Fraunhofer Center Digital Energy, Fraunhofer FIT, Aachen, Germany\\Email: f.klein-helmkamp@iaew.rwth-aachen.de
    }}

\maketitle

\begin{abstract}
Flexibility is increasingly gaining importance in modern power system operation. This paper presents a controller framework based on Online Feedback Optimization for real-time coordination of power system flexibility. The proposed approach introduces a momentum-augmented projection-step to accelerate convergence and improve dynamic performance. We derive the controller formulation, and evaluate its performance and stability in two representative case studies. The first examines online congestion management in distribution feeders, and the second addresses multi-layer flexibility dispatch across system interfaces. Numerical results demonstrate that the momentum-based controller achieves faster convergence and maintains constraint satisfaction, highlighting its potential for real-time flexibility control in large-scale power systems.
\end{abstract}


\section{Introduction}
Modern power systems are increasingly operated near their technical limits with reduced stability margins, heightening the need for operational flexibility to maintain secure and reliable operation \cite{entsoe_2024}. Distributed energy resources (DERs), such as wind power plants, PV systems, and battery energy storage systems, offer significant potential to provide this flexibility \cite{Impram_2020}. However, their distribution in the underlying sub-transmission systems poses technical challenges for the coordination of ancillary services \cite{Kryonidis_2021}. These include uncertainty due to volatility, computational complexity, and model inaccuracy for control and optimization. Online Feedback Optimization (OFO) has recently gained attention for different use cases in the operation of power systems \cite{Hauswirth_2024, Ortmann_2023}. Since OFO integrates optimization algorithms directly into a closed-loop control framework, careful controller tuning is essential to ensure stability and robustness in flexibility coordination \cite{Zettl_2024,Klein-Helmkamp_2025}.
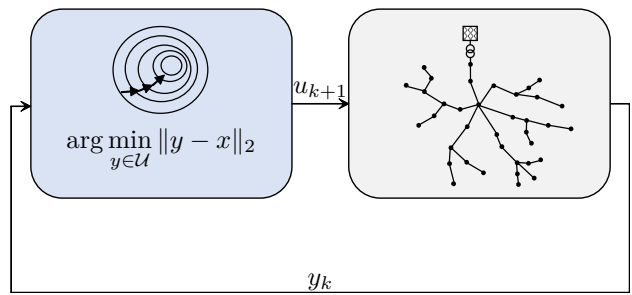
\begin{figure}[tb]
	\centering
	%% Creator: Inkscape 1.3.2 (091e20e, 2023-11-25, custom), www.inkscape.org
%% Accompanies image file 'controller.pdf' (pdf, eps, ps)
\begingroup%
  \makeatletter%
  \providecommand\color[2][]{%
    \errmessage{(Inkscape) Color is used for the text in Inkscape, but the package 'color.sty' is not loaded}%
    \renewcommand\color[2][]{}%
  }%
  \providecommand\transparent[1]{%
    \errmessage{(Inkscape) Transparency is used (non-zero) for the text in Inkscape, but the package 'transparent.sty' is not loaded}%
    \renewcommand\transparent[1]{}%
  }%
  \providecommand\rotatebox[2]{#2}%
  \newcommand*\fsize{\dimexpr\f@size pt\relax}%
  \newcommand*\lineheight[1]{\fontsize{\fsize}{#1\fsize}\selectfont}%
  \ifx\svgwidth\undefined%
    \setlength{\unitlength}{237.59999657bp}%
    \ifx\svgscale\undefined%
      \relax%
    \else%
      \setlength{\unitlength}{\unitlength * \real{\svgscale}}%
    \fi%
  \else%
    \setlength{\unitlength}{\svgwidth}%
  \fi%
  \global\let\svgwidth\undefined%
  \global\let\svgscale\undefined%
  \makeatother%
  \begin{picture}(1,0.4848485)%
    \lineheight{1}%
    \setlength\tabcolsep{0pt}%
    \put(0,0){\includegraphics[width=\unitlength,page=1]{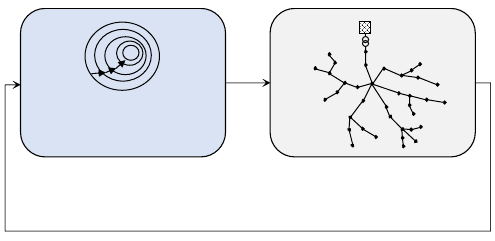}}%
    \put(0.50080982,0.33463004){\color[rgb]{0,0,0}\makebox(0,0)[t]{\lineheight{1.25}\smash{\begin{tabular}[t]{c}$u_{k+1}$\end{tabular}}}}%
    \put(0.50064598,0.03058968){\color[rgb]{0,0,0}\makebox(0,0)[t]{\lineheight{1.25}\smash{\begin{tabular}[t]{c}$y_{k}$\end{tabular}}}}%
    \put(0.24853554,0.2466488){\color[rgb]{0,0,0}\makebox(0,0)[t]{\lineheight{1.25}\smash{\begin{tabular}[t]{c}$\arg\min_{y \in \mathcal{U}} \|y - x \|_2$\end{tabular}}}}%
  \end{picture}%
\endgroup%

	\caption{Closed-loop implementation of projected gradient descent.}
	\label{fig:pgd}
\end{figure}

\subsection{Related Work}
Flexibility has become an increasingly studied topic in the operation of modern power systems. While specific use cases differ, the activation of flexibility generally requires efficient coordination of numerous and spatially distributed actuators within the system \cite{Früh_2023}. A commonly adopted approach to determine the dispatch of available flexibility is to formulate the task as an optimization problem. These problems can be solved centrally or decomposed when flexibility must be coordinated across system boundaries \cite{Givisiez_2020}. In order to ensure that the dispatch of flexibility is fulfilled subject to the limitations of the system, a grid model in form the power flow equations needs to be included in the optimization problem. Consequently, the dispatch problem typically takes the form of an AC Optimal Power Flow (ACOPF). Because the ACOPF is computationally challenging, various methods have been proposed to make the problem more tractable \cite{Molzahn_2019}. The authors of \cite{Torbaghan_2020} propose a relaxation of the power flow equations using second-order cones. Other approaches rely on the DistFlow or LinDistFlow approximations to improve tractability for large-scale distribution systems. For example, \cite{Bandeira_2024} presents a method to aggregate flexibility at the interface between systems using the DistFlow equations, while \cite{Prat_2024} employs LinDistFlow to ensure network constraint satisfaction, further simplifying the nonlinear and nonconvex power flow equations. The aforementioned works solve the dispatch problem in closed form through feedforward optimization. Although this approach is common in power systems, it can yield suboptimal results in the presence of disturbances and uncertainty, and it remains computationally demanding due to the nonconvex nature of the ACOPF problem \cite{Narimani_2018}. To address these challenges, \cite{Escobar_2025} proposes a data-driven method enabling sensitivity-based participation of DERs in transmission-level voltage control. The use of such sensitivities facilitates closed-loop optimization methods, such as Model Predictive Control (MPC) or, more recently, Online Feedback Optimization \cite{Colombino_2018, Picallo_2022}. OFO is particularly advantageous in cases where system dynamics can be approximated by their steady-state behavior \cite{Häberle_2020}, as it relies on sensitivities that are computationally inexpensive and does not require detailed system models \cite{Ortmann_2024}. This characteristic is especially relevant for applications operating on time scales slower than fast frequency and voltage dynamics, including real-time congestion management, voltage control in distribution feeders, and reference tracking at interfaces between grid layers. Several studies have applied OFO in power system operations. In \cite{Bernstein_2019}, a feedback-optimization framework is proposed for the provision of requested services from DERs using primal–dual projected gradient methods. The authors address the tracking of time-varying optimization problems in closed loop to ensure constraint satisfaction. In \cite{Zhan_2024}, an OFO-based control strategy for DERs connected to distribution grids is introduced, combining superimposed optimization for intertemporal planning with real-time control. Building on this, \cite{Zhan_2025} presents a framework for active distribution grid control that implements an OFO controller via a primal–dual gradient projection method, systematically evaluating its robustness under adverse conditions and introducing metrics to quantify the impact of disturbances on controller performance. The works presented so far only consider standard implementations of gradient-projection-based optimization algorithms without added momentum. This is common practice, e.g. in machine learning \cite{Qian_1999, Dozat_2016}, but is not often considered in real-time control applications. In \cite{Hauswirth_2019}, a theoretical bound for the stability of momentum-based optimization dynamics was introduced, such as accelerated gradient descent in closed-loop configurations. The authors highlight the potential to increase convergence speed but also discuss the challenge in guaranteeing stability for variants that do not exhibit uniform asymptotic convergence.

As shown through the literature review OFO has been considered for a variety of interesting use cases in power systems operations. Most implementations focus on standard variants of projected gradient descent (PGD), primal dual gradient projection, or saddle flows. The inclusion and effect of momentum-terms in online optimization of power systems remains underexplored.

\subsection{Main contribution}
To address the identified research gap we propose a cascaded OFO controller to track set points for power flows at the interface between system operators such as \cite{Klein-Helmkamp_2024}. As our main contribution, we introduce a momentum-augmented feedback law that exploits both current and past gradient information within the projection step to accelerate convergence and improve stability. Our paper investigates the following research questions:
\begin{enumerate}
    \item What is the impact of the momentum term on the dynamic behavior of the proposed controller?
    \item How can the controller be tuned to ensure stable and efficient flexibility dispatch in online power system operation?
\end{enumerate}
Building on existing theoretical guarantees for momentum-based OFO, we analyze the closed-loop behavior of the proposed controller and demonstrate that it achieves stability and faster convergence in realistic grid scenarios. These properties make it well suited for real-time applications such as distributed flexibility coordination. We evaluate the performance of our approach in two numerical case studies for different use cases. First, we investigate a scenario of online congestion management in distribution feeders. Second, we evaluate the controllers ability to track set points for power flow at the coupling transformer as part of a multi-layer flexibility dispatch scheme.

\section{Controller Framework}
\label{sec:controller}
\subsection{General Optimization Problem}
Operational tasks in power systems, such as generation dispatch, voltage regulation, and flexibility coordination, are expressed most naturally as constrained optimization problems. In such formulations, system-level objectives are represented by cost functions, while network and actuator limitations appear as explicit constraints. This structure provides a principled foundation for coordinating multiple controllable resources (such as DERs and loads) to achieve technically efficient operating points. We describe the objectives of the grid operator in our framework as a constrained optimization problem of the form:
\begin{equation}
\label{eq:opt_general}
\begin{aligned}
    \min_{x, u} & \Phi (x,u) \\
    \text{s.t.} \quad
    & x = h(u) \\
    & x \in \mathcal{X} \\
    & u \in \mathcal{U}
\end{aligned}
\end{equation}
where $\Phi : \mathbb{R}^{n} \times \mathbb{R}^m \rightarrow \mathbb{R}$ denotes the system's cost function, $u \in \mathbb{R}^m$ the vector of controllable inputs, and $x\in \mathbb{R}^n$ the vector of system states (e.g. voltages and power flows). The feasible input set $\mathcal{U} \subset \mathbb{R}^m$ denotes actuator constraints. While the feasible output set $\mathcal{X} \subset \mathbb{R}^n$ denotes constraints such as voltage bands and thermal limits. We denote the minimizer of \eqref{eq:opt_general} with $u^{\star}$. The vector $x \in \mathbb{R}^n$ denotes the vector of system states, e.g. bus voltages and power flow. The mapping $h: \mathbb{R}^m \rightarrow \mathbb{R}^n$ captures the steady-state relationship between control inputs and system states. In the case of AC power systems, $h(u)$ is implicitly defined by the nonlinear AC power flow equations, which relate bus voltages to active and reactive power injections through Kirchhoff’s laws and network admittances for all system busses $i \in \mathcal{N}$. That is, the steady-state voltages $v \in \mathbb{C}^n$ satisfy
\begin{equation}
    s_i = v_i \sum_{j \in \mathcal{N}} y_{ij}^* v_j^*, \qquad \forall i \in \mathcal{N},
\end{equation}
where $s_i = p_i + jq_i$ represents the complex power injection at bus $i$, and $y_{ij}$ denotes the $(i,j)$-th entry of the network admittance matrix $Y$. The mapping $h(\cdot)$ thus acts as a static nonlinear system that links the controllable inputs $u$ to the states $x$. The optimization problem \eqref{eq:opt_general} thus defines the desired steady-state operating point of the physical system. In closed-loop operation, this formulation serves as the target for the proposed feedback optimization controller, which drives the system toward the minimizer of~\eqref{eq:opt_general} while maintaining feasibility with respect to $x=h(u)$, $x\in \mathcal{X}$, and $u \in \mathcal{U}$. In the following section, we detail the specific formulation used in this work.

\subsection{Online Feedback Optimization}
Given the constrained optimization problem \eqref{eq:opt_general} that encodes the control objective of the system operator, the proposed OFO controller is implemented in closed loop through the following sequence of steps. One full execution corresponds to a single OFO iteration:
\begin{enumerate}
	\item \emph{State acquisition}: Acquire the current state of the grid $y$ by directly measuring the relevant state of the system.
	\item \emph{Gradient update}: Update the gradient of the cost function $\nabla \Phi(u,y)$ for the current step taking into account the measured states.
	\item \emph{Projection step}: Determine the next control action by projecting the gradient onto the feasible set $\mathcal{U}$ and scaling the solution with $\alpha$.
	\item \emph{Actuation of system}: Send determined set points to available actuators within the system under control.
\end{enumerate}
We linearize the nonlinear mapping \(x = h(u)\), defined by the AC power flow equations, around a nominal operating point \((x_0, u_0)\) that satisfies the steady-state equations. We approximate small variations in the system states and control inputs as
\begin{equation}
    \Delta y = \nabla h\, \Delta u,
\end{equation}
where \(\nabla h = \sfrac{\partial h}{\partial u}\big|_{(x_0,u_0)}\) denotes the sensitivity matrix of the network. This linearization allows us to express the system output as an affine function of the control input. For the vector of state measurements $y\in \mathbb{R}^n$ we write the approximated system output with
\begin{equation}
    y \approx y_0 + \nabla h\,(u - u_0).
\end{equation}
The sensitivity matrix \(\nabla h(u)\) captures the local geometry of the AC power flow manifold by describing how infinitesimal variations in controllable injections \(u\) affect the corresponding system outputs such as voltages and other state variables \cite{Bolognani_2015}. We use this linearized relation to encode the network constraints in the controller design and to formulate the projection step as a quadratic program. Specifically, by incorporating the sensitivities \(\nabla h(u)\) and the gradient information of the objective function \(\Phi(u,y)\), we obtain the quadratic problem (QP) used to project the gradient $\nabla\Phi$ onto the feasible set $\mathcal{U}$. We denote the matrix $C \in \mathbb{R}^{n_c \times m}$ as the input constraint matrix and matrix $D \in \mathbb{R}^{n_d \times n}$ as the output constraint matrix of the system and write the QP with 
\begin{equation}
	\label{eq:ofo_qp}
    \begin{aligned}
		\sigma = \arg \min_{w \in \mathbb{R}^m} 
		& \|w + G^{-1}H(u)^T \nabla \Phi(u, y) \|^2 \\
		\text{s.t.} \quad 
		& C (u(k) + \alpha w) \leq b, \\
		& D (y(k) + \alpha \nabla h(u) w) \leq d.
	\end{aligned}
\end{equation}
computing the direction \(\sigma\) that projects the controller update onto the feasible region defined by the constraints of the system. The relative importance of control adjustments is weighted in the cost function with the positive definite matrix $G \succ 0$. We subsequently compute the input vector for step $k+1$ by scaling the solution of \eqref{eq:ofo_qp} with the gain parameter $\alpha \ \in (0,1]$ before adding it to $u_{k}$ with:
\begin{equation}
    u_{k+1} = u_{k} + \alpha \sigma
\end{equation}
The gain $\alpha$ represents an important tuning parameter in both standard variants and momentum-accelerated versions of PGD as it bounds the system output and influences convergence speed. To ensure stability of both optimization and system dynamics, $\alpha$ needs to be chosen appropriately small, potentially reducing the convergence speed and therefore the controller performance. To potentially counteract this, we propose to introduce momentum to PGD in OFO. We describe our specific implementation in the following subsection.

\subsection{Momentum-Accelerated Projected Gradient Descent}
We extend the projected gradient-based controller by introducing a momentum term that exploits information from the previous iteration to accelerate convergence. Before performing the projection step, we modify the descent direction by combining the current and previous gradients of the cost function. This modification introduces a single-step memory into the controller, effectively adding an inertial component to the optimization dynamics. We apply the momentum step before the projection to ensure that the subsequent projection onto the feasible set still guarantees satisfaction of the operational and physical constraints. At iteration \(k\), we compute the momentum-augmented gradient as
\begin{equation}
\label{eq:momentum}
\nabla \tilde{\Phi}_k 
= \beta\, \nabla \Phi_k + (1 - \beta)\, \nabla \Phi_{k-1},
\qquad \beta \in (0,1],
\end{equation}
where \(\nabla \Phi_k := \nabla_u \Phi(u(k), y(k))\) denotes the current gradient and \(\nabla \Phi_{k-1}\) the one obtained at the previous time step. The scalar \(\beta\) weights the relative contribution of the two terms: a value close to one recovers the standard projected gradient step, while smaller values increase the influence of the previous gradient. It therefore serves as an additional tuning parameter for the controller behavior.  

\section{Case Study I: Congestion Management}
\begin{figure}[tb]
	\centering
	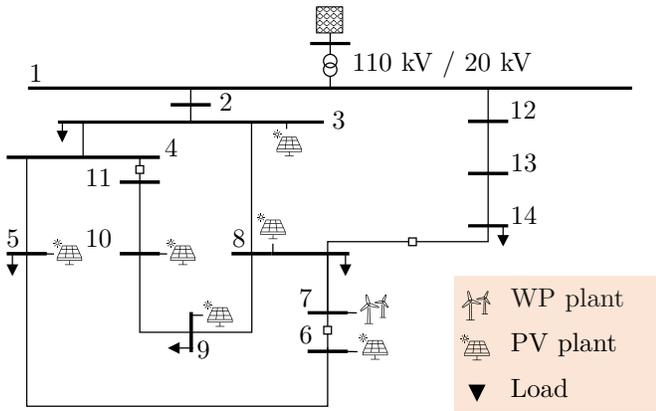
	\caption{Used configuration of CIGRE MV benchmark system \cite{cigre_2014}. Open switches are indicated by boxes.}
	\label{fig:mv_benchmark}
\end{figure}

We first demonstrate the proposed OFO controller for the application of online congestion management in distribution grids. In this use case, the controller monitors the grid state in real time and activates distributed flexibility to alleviate operational limit violations. The CIGRE MV benchmark network \cite{cigre_2014}, shown in \autoref{fig:mv_benchmark}, serves as the test system. The study focuses on the left feeder beginning at bus~2.
\subsection{Problem Setup}
The objective is to adjust the active and reactive power setpoints of distributed energy resources (DERs) such that all operational constraints are satisfied while minimizing deviation from the nominal dispatch schedule. This formulation captures the trade-off between maintaining planned operation and ensuring secure network conditions.  We introduce a quadratic cost function $\Phi$ that penalizes deviations of the active power injections from their scheduled values, while also discouraging excessive reactive power usage. Let $p \in \mathbb{R}^m$ and $q \in \mathbb{R}^m$ denote the vectors of active and reactive power injections, respectively, while $p^{\mathrm{nom}} \in \mathbb{R}^m$ represents the nominal active power reference for the given time step. The weighting matrices $A, B \succeq 0$ define the relative importance of active and reactive power deviations in the cost function. The optimization problem implemented in the OFO controller is expressed as
\begin{equation}
	\label{eq:voltage_control_problem}
	\begin{aligned}
		\min_{p,q} &\quad \Phi
= (p - p^{\mathrm{nom}})^\top A \,(p - p^{\mathrm{nom}})
\;+\; q^\top B\, q, \\
		\text{s.t.}
		&\quad v_{\text{min}} \leq v
			\leq v_{\text{max}}, \\
		&\quad s_{\text{min}} \leq s
			\leq s_{\text{max}}, \\
		&\quad p_{\text{min}} \leq p \leq p_{\text{max}}, \\
		&\quad q_{\text{min}} \leq q \leq q_{\text{max}}.
	\end{aligned}
\end{equation}
The output constraints are enforced using real-time measurements of bus voltage magnitudes $v \in \mathbb{R}^{n_v}$ and branch flows $s \in \mathbb{R}^{n_s}$, which are compared to their respective limits at each iteration. In this formulation, the nonlinear AC power flow equations are not explicitly included in the optimization problem; instead, their effect is captured implicitly through the measurement feedback. This structure enhances tractability and makes the approach suitable for real-time implementation, in contrast to conventional ACOPF formulations. To solve \eqref{eq:voltage_control_problem} in closed-loop we embed it into the momentum-accelerated formulation of PGD introduced in \autoref{sec:controller}. The capabilities of DERs and loads are integrated into the input constraint matrix $C$ and voltage and load flow constraints are written into the output constraint matrix $D$.

\subsection{Results and Analysis}
\begin{figure}[b]
	\centering
	\input{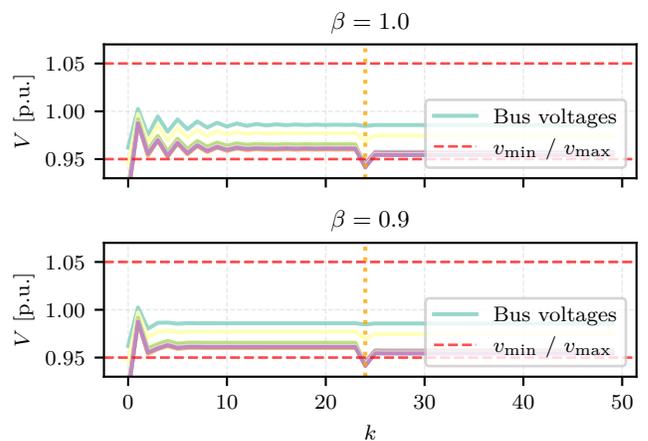}
    \vspace{-0.8cm}
	\caption{Bus voltages during online congestion management with generation disconnection event at $k=24$.}
	\label{fig:case_1_voltage}
\end{figure}

\begin{figure}[tb]
	\centering
	\input{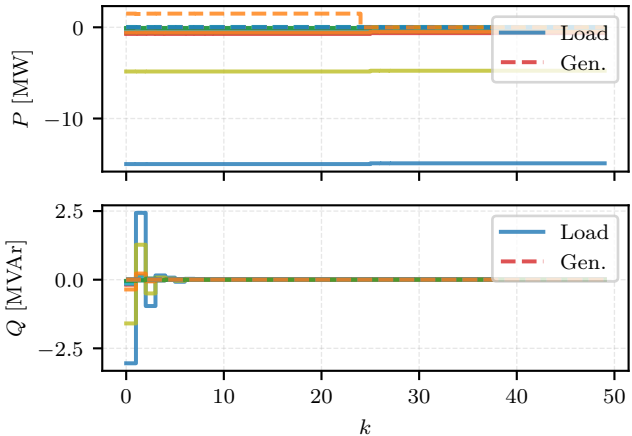}
    \vspace{-1cm}
	\caption{In-feed and consumption of active and reactive power of flexible loads and generators for $\beta=0.9$.}
	\label{fig:case_1_actors}
\end{figure}

To evaluate the performance of the proposed OFO controller for online congestion management, we initialize the system with a voltage constraint violation ($v_n \pm 5\%$) and apply the optimization problem defined in \eqref{eq:voltage_control_problem}. Additionally, we set reactive power operating points for load and generation to demonstrate how the controller prioritizes active and reactive power differently. To further assess the robustness of the controller to disturbances, we simulate the sudden disconnection of the wind power plant (WPP) at bus~7 with a rated power of $p_{\text{n}} = 1.5~\text{MW}$. \autoref{fig:case_1_voltage} illustrates the evolution of bus voltages for different momentum parameters $\beta \in [0.9, 1]$. The controller gain is chosen consistently with $\alpha=0.8$.
The results show that the initial voltage violation is corrected within a single iteration for both controller configurations. This follows from the projection step, which ensures that each controller update remains within the feasible region of problem~\eqref{eq:voltage_control_problem}. After the constraint is resolved, the system settles into a steady state until the WPP disconnection at $k=24$ causes a temporary undervoltage. The controller again restores all voltages to their admissible range within one iteration, demonstrating its ability to react promptly to sudden disturbances. Introducing momentum ($\beta=0.9$) noticeably improves transient behavior. The oscillations primarily originate from the coordinated adjustments of the flexible resources. \autoref{fig:case_1_actors} depicts the evolution of active and reactive power injections for the flexible units. The controller predominantly exploits reactive power to regulate voltages, while maintaining active power close to its nominal reference, consistent with the cost function of \eqref{eq:voltage_control_problem}. Overall, through the use of momentum, we can achieve stable operation and fast convergence to a steady state even under high gain values $\alpha$.
 
\section{Case Study II: Flexibility Dispatch}
Building on the results of the previous case study, we now evaluate the proposed OFO controller for tracking a power flow reference at the system interface between grid layers. Considering the CIGRE MV benchmark system, we explicitly include two individual underlying LV systems at bus~5 and bus~8, both coordinated by individual OFO controllers \cite{Meinecke_2020}. This extends the single layer controller to include small-scale flexibility in further underlying grids.

\subsection{Problem Setup}
\begin{figure}[tb]
\centering
\begin{tikzpicture}[
    controller/.style={rectangle, draw, rounded corners, thick, 
                       minimum width=2cm, minimum height=0.5cm,
                       align=center, fill=blue!10},
    arrowdown/.style={-{Latex[length=3mm]}, thick},
    arrowup/.style={-{Latex[length=3mm]}, thick, dashed},
    node distance=0.8cm and 1cm
]

% Nodes
\node[controller] (C1) {HV grid};
\node[controller, below=of C1] (C2) {MV grid};
\node[controller, below right=of C2] (C3) {LV grid 1};
\node[controller, below left=of C2] (C4) {LV grid 2};

% Downward setpoint arrows
\draw[arrowdown] (C1) to[bend left=15] node[midway, right] {$s_{\text{set}}$} (C2);
\draw[arrowdown] (C2) to[bend left=15] node[midway, below right] {$s_{\text{set}}$} (C4);
\draw[arrowdown] (C2) to[bend right=15] node[midway, below left] {$s_{\text{set}}$} (C3);

% Upward return arrows (feedback)
\draw[arrowup] (C2) to[bend left=15] node[midway, left] {$s_{\text{PCC,HV-MV}}$} (C1);
\draw[arrowup] (C3) to[bend right=15] node[midway, above right] {$s_{\text{PCC,MV-LV1}}$} (C2);
\draw[arrowup] (C4) to[bend left=15] node[midway, above left] {$s_{\text{PCC,MV-LV2}}$} (C2);

\end{tikzpicture}
\caption{Hierarchical structure of interacting OFO controllers.}
\label{fig:case_2_setup}
\end{figure}
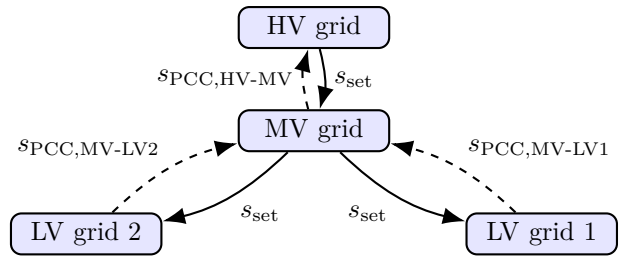

We consider the vertical flexibility dispatch problem illustrated in \autoref{fig:case_2_setup}. The objective of the grid operator is to track a reference power flow at the point of common coupling (PCC) between adjacent grid layers. In this context, the activation of flexibility corresponds to a shift in the steady-state operating point at the coupling transformer. To enable coordination across grid layers without exchanging detailed models or customer-specific data, we decompose the optimization problem along the system boundaries, i.e., at each PCC. This formulation supports hierarchical interaction between system operators in a cascaded structure (e.g. EHV to HV system to multiple underlying MV and LV grids) while preserving the autonomy of underlying networks. Flexibility requests are assumed to propagate downstream, where each subordinate layer adjusts its internal operation to fulfill the active and reactive power targets received from its upstream controller. The control objective for a given grid layer is to minimize the Euclidean distance between the actual and requested power exchange at the PCC. We formalize the problem as:
\begin{equation}
	\label{eq:dispatch problem}
	\begin{aligned}
		\min_{p,q} &\quad \Phi =  ||p_{\text{set}} - p_{\text{PCC}}||^{2} + ||q_{\text{set}} - q_{\text{PCC}}||^{2}  \\
		\text{s.t.}
		&\quad v_{\text{min}} \leq v \leq v_{\text{max}}, \\
		&\quad s_{\text{min}} \leq s \leq s_{\text{max}}, \\
		&\quad p_{\text{min}} \leq p \leq p_{\text{max}}, \\
		&\quad q_{\text{min}} \leq q \leq q_{\text{max}}
	\end{aligned}
\end{equation}
Here, $p_{\text{PCC}}$ and $q_{\text{PCC}}$ denote the measured active and reactive power flows at the interface to the upper grid layer, while $p_{\text{set}}$ and $q_{\text{set}}$ represent the corresponding reference values communicated from the upstream controller. Voltage magnitude and thermal loading constraints are enforced through the bounds on $v$ and $s$ respectively. Limits on available flexibility from local resources and subordinate networks are incorporated directly into the input constraints, analogous to generator and load capability limits. Matrices $C$ and $D$ encode these input and output constraints for inclusion in the projection step of the OFO formulation \eqref{eq:ofo_qp}. This ensures that the controller tracks the desired power flow at the PCC without violating internal operational constraints of the participating system.

\subsection{Results and Analysis}
\begin{figure}[tb]
	\centering
	\input{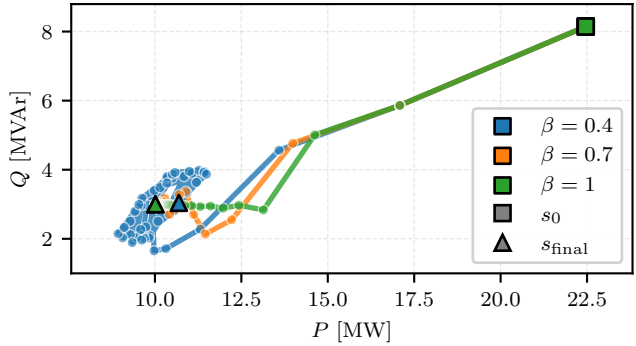}
    \vspace{-1cm}
	\caption{P-Q plane trajectories showing convergence behavior for different momentum $\beta$.}
	\label{fig:case_2_pq}
\end{figure}

\begin{figure}[tb]
    \centering
    \input{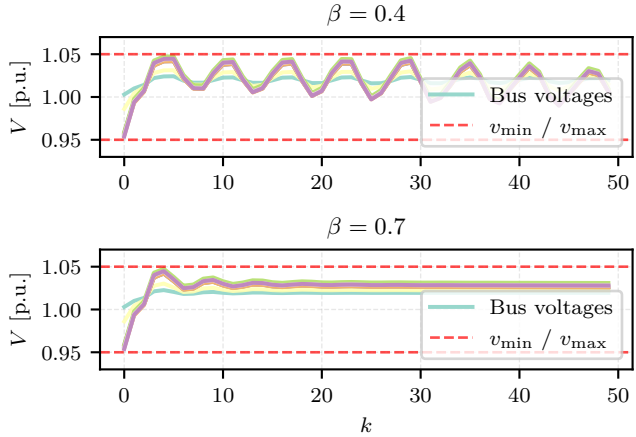}
    \vspace{-1cm}
    \caption{Bus voltages during flexibility provision using different momentum parameters.}
    \label{fig:mv_voltages_comparison}
\end{figure}

\begin{figure}[t]
	\centering
	\input{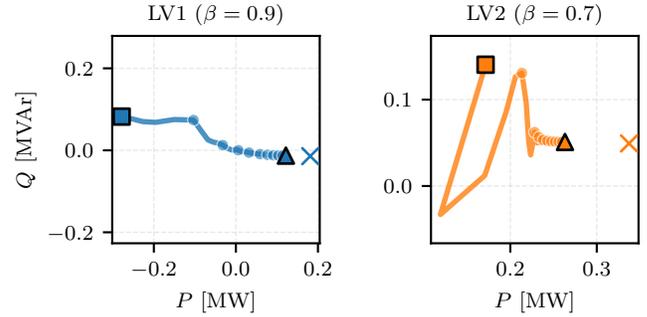}
    \vspace{-1cm}
	\caption{P-Q planes of MV-LV transformers showing convergence of trajectories to provide requested flexibility to superimposed grid.}
	\label{fig:case_2_pq_lv}
\end{figure}

\begin{figure}[b]
	\centering
	\input{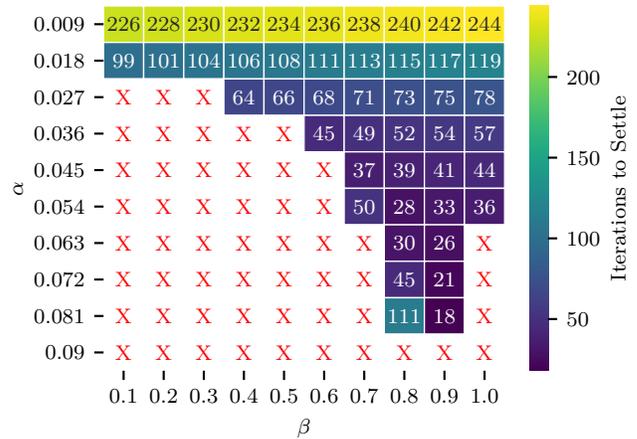}
    \vspace{-0.8cm}
	\caption{Iterations to settle for varying $\alpha$ and $\beta$ parameters. Non-converged trajectories are marked with a red X.}
	\label{fig:case2_alpha_beta}
\end{figure}

We analyze the convergence behavior of the proposed OFO controller for different momentum parameters $\beta \in [0.4,0.7,1]$ when tracking the same reference power flow at the system interface. All trajectories thus converge toward the same minimizer $u^{\star}$ of the optimization problem \eqref{eq:dispatch problem}. The goal of this analysis is to characterize how the choice of $\beta$ influences the optimization dynamics. For this experiment, the controller gain is fixed to $\alpha=0.05$, and the reference set point is $p_{\text{set}}, q_{\text{set}} = (10~\text{MW}, 3~\text{MVAr})$. The resulting trajectories on the P-Q-plane of the HV-MV-transformer are depicted in \autoref{fig:case_2_pq}. From the second iteration onward, the influence of the momentum term becomes evident, as the previous-step gradient is incorporated into the update. Increasing momentum decreases the needed iterations to reach the requested set point. The triangles in the figure marks the final state of the trajectory. The corresponding voltage magnitudes during flexibility provision are shown in \autoref{fig:mv_voltages_comparison}.
At the lower grid levels, the underlying LV systems adjust their operation in response to flexibility requests from the superimposed OFO controller. The resulting power trajectories at the MV–LV transformers are shown in \autoref{fig:case_2_pq_lv}. Both LV systems exhibit convergence toward the requested set point (marked by an X), though the exact target is not fully reached within the simulated iterations. This deviation is caused by active system constraints, which limit the feasible operating region. The hierarchical decomposition thus ensures that local constraints are automatically respected--an inherent advantage of the proposed OFO cascade.
To further evaluate the sensitivity of convergence speed and stability to the controller parameters, we conduct a parameter sweep over $\alpha$ and $\beta$. \autoref{fig:case2_alpha_beta} shows the number of iterations required for convergence to $p_{\text{set}}, q_{\text{set}} = (10~\text{MW}, 3~\text{MVAr})$. Non-convergent trajectories within 300 iterations are marked with red crosses. The heat map reveals a clear dependence of convergence speed on both parameters. Increasing the gain from $\alpha = 0.009$ to $\alpha = 0.054$ reduces the number of iterations from $k=244$ to $k=36$, significantly improving performance without compromising stability. A similar effect is observed for increasing momentum. For a fixed gain, lower $\beta$ values generally accelerate convergence but may induce oscillations near the stability boundary, e.g. $(\alpha , \beta) = (0.054, 0.7) \rightarrow k=50$ and $(\alpha , \beta) = (0.081, 0.8) \rightarrow k=111$. The results further show a narrowing of the stability region as $\alpha$ increases, with instability occurring at $\beta=1$ for $\alpha \geq 0.063$. In this regime, slightly reducing $\beta$ restores stability by providing additional damping. The fastest stable convergence is achieved at $(\alpha,\beta) =(0.081, 0.9)$, requiring only $k=18$ iterations. 
Overall, the results demonstrate that incorporating momentum into the OFO controller substantially enhances convergence performance while maintaining stable operation across multiple grid layers. The empirical analysis confirms that appropriate tuning of $\alpha$ and $\beta$ allows for efficient flexibility dispatch in hierarchical, real-time coordination settings.

\section{Conclusion}
We presented an OFO controller for real-time coordination of distributed flexibility in power systems. The proposed method integrates a momentum-term into a sensitivity-based PGD framework. By averaging the current and previous gradients before the projection step, the controller achieves smoother and faster convergence while maintaining feasibility through a quadratic-program-based update law. Two case studies based on the CIGRE MV benchmark validated the approach. In an online congestion-management scenario, the controller eliminated voltage-band violations within a single iteration, prioritized reactive power corrections to preserve active power schedules, and remained robust under disturbance. In the hierarchical setpoint-tracking case, the inclusion of the momentum parameter~$\beta$ effectively reduced the number of iterations required to reach the desired $P$–$Q$ targets at system interfaces, all while respecting the network constraints in the MV grid and in the subordinate LV systems. We empirically observed that simultaneous tuning of $\alpha$ and $\beta$ leads to a generally tighter stability region of the controller. Overall, the results demonstrate that introducing momentum into OFO can enhance convergence speed for real-time flexibility coordination in power systems and is a viable approach for use cases that require fast and secure action, even with disturbances and model uncertainties present.

\section*{Acknowledgment}
\begin{wrapfigure}{r}{0.12\textwidth}
  \vspace{-\baselineskip}
  \vspace{-\baselineskip}
  %\centering
  \includegraphics[width=0.12\textwidth]{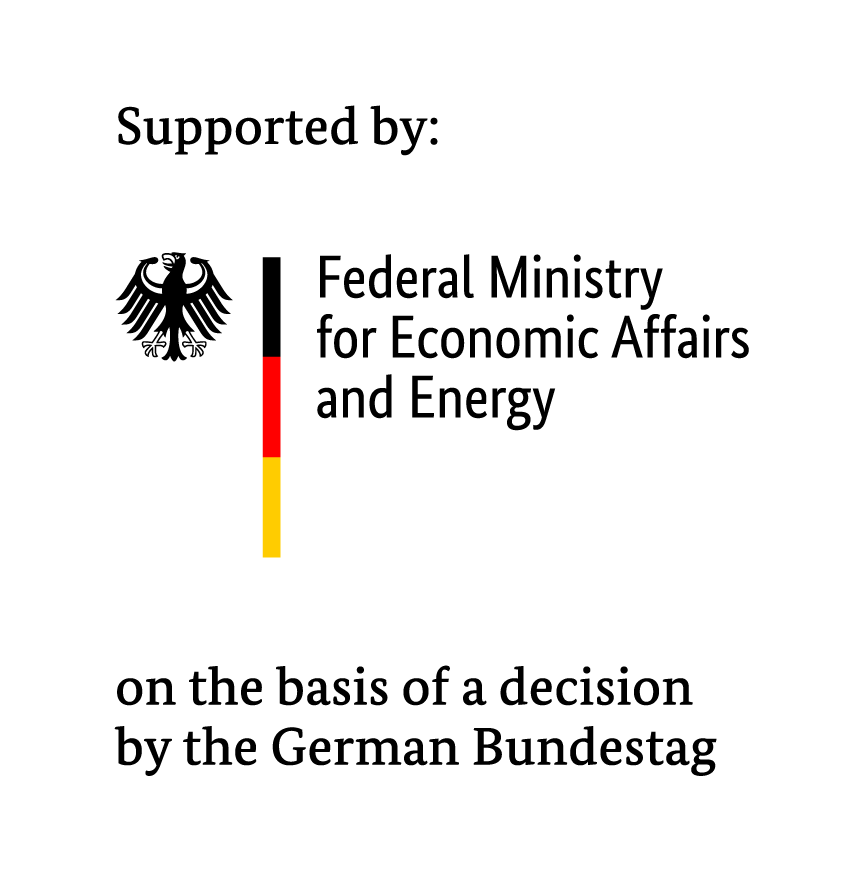}
  \vspace{-\baselineskip}
    \vspace{-\baselineskip}
        \vspace{-\baselineskip}
\end{wrapfigure}

This project received funding from the German Federal Ministry for Economic Affairs and Climate Action under the agreement no. 03EI4046E (PROGRESS).


\begin{thebibliography}{00}

\bibitem{entsoe_2024}
entso-e, 2024. System Flexibility Needs for the Energy Transition. entso-e, Brussels, Belgium.

\bibitem{Impram_2020}
Impram, S., Varbak Nese, S., Oral, B., 2020. Challenges of renewable energy penetration on power system flexibility: A survey. Energy Strategy Reviews 31, 100539.

\bibitem{Kryonidis_2021}
Kryonidis, G.C., Kontis, E.O., Papadopoulos, T.A., Pippi, K.D., Nousdilis, A.I., Barzegkar-Ntovom, G.A., Boubaris, A.D., Papanikolaou, N.P., 2021. Ancillary services in active distribution networks: A review of technological trends from operational and online analysis perspective. Renewable and Sustainable Energy Reviews 147, 111198.

\bibitem{Hauswirth_2024}
Hauswirth, A., He, Z., Bolognani, S., Hug, G., Dörfler, F., 2024. Optimization algorithms as robust feedback controllers. Annual Reviews in Control 57, 100941.

\bibitem{Ortmann_2023}
Ortmann, L., Hotz, G., Bolognani, S., Dörfler, F., 2023. Real-time Curative Actions for Power Systems via Online Feedback Optimization, in: 2023 IEEE Belgrade PowerTech,  IEEE, Belgrade, Serbia, pp. 1–6.

\bibitem{Zettl_2024}
Zettl, I., Klein-Helmkamp, F., Schmidtke, F., Ortmann, L., Ulbig, A., 2024. Tuning a Cascaded Online Feedback Optimization Controller for Provision of Distributed Flexibility, in: 2024 International Conference on Smart Energy Systems and Technologies (SEST), IEEE, Torino, Italy, pp. 1–6.

\bibitem{Klein-Helmkamp_2025}
Klein-Helmkamp, F., Möllemann, T., Zettl, I., Ulbig, A., 2025. Stability and Performance of Online Feedback Optimization for Distribution Grid Flexibility, in: 2025 IEEE Kiel PowerTech. Presented at the 2025 IEEE Kiel PowerTech, IEEE, Kiel, Germany, pp. 1–6.

\bibitem{Früh_2023}
Früh, H., Müller, S., Contreras, D., Rudion, K., Von Haken, A., Surmann, B., 2023. Coordinated Vertical Provision of Flexibility Freom Distribution Systems. IEEE Trans. Power Syst. 38, 1834–1844.

\bibitem{Givisiez_2020}
Givisiez, A.G., Petrou, K., Ochoa, L.F., 2020. A Review on TSO-DSO Coordination Models and Solution Techniques. Electric Power Systems Research 189, 106659.

\bibitem{Molzahn_2019}
Molzahn, D.K., Hiskens, I.A., 2019. A Survey of Relaxations and Approximations of the Power Flow Equations. FNT in Electric Energy Systems 4, 1–221.    

\bibitem{Torbaghan_2020}
Torbaghan, S.S., Suryanarayana, G., Hoschle, H., D’hulst, R., Geth, F., Caerts, C., Van Hertem, D., 2020. Optimal Flexibility Dispatch Problem Using Second-Order Cone Relaxation of AC Power Flows. IEEE Trans. Power Syst. 35, 98–108.

\bibitem{Bandeira_2024}
Beraldo Bandeira, M., Faulwasser, T., Engelmann, A., 2024. An ADP framework for flexibility and cost aggregation: Guarantees and open problems. Electric Power Systems Research 234, 110818.

\bibitem{Prat_2024}
Prat, E., Dukovska, I., Nellikkath, R., Thoma, M., Herre, L., Chatzivasileiadis, S., 2024. Network-Aware Flexibility Requests for Distribution-Level Flexibility Markets. IEEE Trans. Power Syst. 39, 2641–2652.

\bibitem{Narimani_2018}
Narimani, M.R., Molzahn, D.K., Wu, D., Crow, M.L., 2018. Empirical Investigation of Non-Convexities in Optimal Power Flow Problems, in: 2018 Annual American Control Conference (ACC). Presented at the 2018 Annual American Control Conference (ACC), IEEE, Milwaukee, WI, USA, pp. 3847–3854.

\bibitem{Escobar_2025}
Escobar, F., Pierrou, G., Valverde, G., Hug, G., 2025. Data-driven Participation of Active Distribution Networks in Transmission Voltage Control. Sustainable Energy, Grids and Networks.

\bibitem{Colombino_2018}
Colombino, M., Dall’Anese, E., Bernstein, A., 2020. Online Optimization as a Feedback Controller: Stability and Tracking. IEEE Trans. Control Netw. Syst. 7, 422–432.

\bibitem{Picallo_2022}
Picallo, M., Ortmann, L., Bolognani, S., Dörfler, F., 2022. Adaptive real-time grid operation via Online Feedback Optimization with sensitivity estimation. Electric Power Systems Research 212, 108405.

\bibitem{Häberle_2020}
Häberle, V., Hauswirth, A., Ortmann, L., Bolognani, S., Dörfler, F., 2020. Non-convex Feedback Optimization with Input and Output Constraints. IEEE Control Syst. Lett. 1–1.

\bibitem{Ortmann_2024}
Ortmann, L., Böhm, F., Klein-Helmkamp, F., Ulbig, A., Bolognani, S., Dörfler, F., 2024. Tuning and testing an Online Feedback Optimization controller to provide curative distribution grid flexibility. Electric Power Systems Research 234, 110660.

\bibitem{Bernstein_2019}
Bernstein, A., Dall’Anese, E., 2019. Real-Time Feedback-Based Optimization of Distribution Grids: A Unified Approach. IEEE Trans. Control Netw. Syst. 6, 1197–1209.

\bibitem{Zhan_2024}
Zhan, S., Morren, J., Van Den Akker, W., Van Der Molen, A., Paterakis, N.G., Slootweg, J.G., 2024. Multi-timescale coordinated distributed energy resource control combining local and online feedback optimization. Electric Power Systems Research 234, 110836.

\bibitem{Zhan_2025}
Zhan, S., Morren, J., Van Den Akker, W., Van Der Molen, A., Paterakis, N.G., Slootweg, J.G., 2025. Robustness assessment of primal–dual gradient projection-based online feedback optimization for real-time distribution grid management. Electric Power Systems Research 242, 111468.

\bibitem{Qian_1999}
Qian, N., 1999. On the momentum term in gradient descent learning algorithms. Neural Networks 12, 145–151.

\bibitem{Dozat_2016}
Dozat, T., 2016. Incorporating Nesterov Momentum into Adam, in: Proceedings of the 4th International Conference on Learning Representations. Presented at the International Conference on Learning Representations, pp. 1–4.

\bibitem{Hauswirth_2019}
Hauswirth, A., Bolognani, S., Hug, G., Dorfler, F., 2021. Timescale Separation in Autonomous Optimization. IEEE Trans. Automat. Contr. 66, 611–624.

\bibitem{Klein-Helmkamp_2024}
Klein-Helmkamp, F., Zettl, I., Schmidtke, F., Ortmann, L., Ulbig, A., 2024. Hierarchical provision of distribution grid flexibility with online feedback optimization. Electric Power Systems Research 234, 110779.

\bibitem{Bolognani_2015}
Bolognani, S., Dorfler, F., 2015. Fast power system analysis via implicit linearization of the power flow manifold, in: 2015 53rd Annual Allerton Conference on Communication, Control, and Computing (Allerton). Presented at the 2015 53rd Annual Allerton Conference on Communication, Control and Computing (Allerton), IEEE, Monticello, IL, pp. 402–409.

\bibitem{cigre_2014}
Conseil international des grands réseaux électriques (Ed.), 2014. Benchmark systems for network integration of renewable and distributed energy resources. CIGRÉ, Paris.

\bibitem{Meinecke_2020}
Meinecke, S., Sarajlić, D., Drauz, S.R., Klettke, A., Lauven, L.-P., Rehtanz, C., Moser, A., Braun, M., 2020. SimBench—A Benchmark Dataset of Electric Power Systems to Compare Innovative Solutions Based on Power Flow Analysis. Energies 13, 3290.0


\end{thebibliography}
\end{document}